\documentclass{article}
\usepackage{spconf,amsmath,graphicx}
\usepackage{hyperref}

\usepackage{multirow}
\usepackage{times,booktabs,tabularx,xcolor,url}
\usepackage{algorithm, algpseudocode}
\usepackage{mathtools}
\usepackage{tablefootnote}

\title{T-Foley: A Controllable Waveform-Domain Diffusion Model for Temporal-Event-Guided Foley Sound Synthesis}

\name{$^\sharp$Yoonjin Chung$^*$, $^\sharp$Junwon Lee$^*$, $^\sharp$$^\flat$Juhan Nam 
\thanks{$^*$These authors contributed equally. This work was supported by Institute of Information \& communications Technology Planning \& Evaluation (IITP) grant funded by the Korea government (MSIT) (No.2019-0-00075, Artificial Intelligence Graduate School Program (KAIST)) and the National Research Foundation of Korea (NRF) grant funded by the Korea government (MSIT) (No. RS-2023-00222383).}
}
\address{$^\sharp$Graduate School of Artificial Intelligence, KAIST, Republic of Korea\\
        $^\flat$Graduate School of Culture Technology, KAIST, Republic of Korea}

\begin{document}
\ninept
\maketitle

\begin{abstract}
Foley sound, audio content inserted synchronously with videos, plays a critical role in the user experience of multimedia content. Recently, there has been active research in Foley sound synthesis, leveraging the advancements in deep generative models. However, such works mainly focus on replicating a single sound class or a textual sound description, neglecting temporal information, which is crucial in the practical applications of Foley sound.
We present \textbf{\textit{T-Foley}}, a \textbf{T}emporal-event-guided waveform generation model for \textbf{Foley} sound synthesis. T-Foley generates high-quality audio using two conditions: the sound class and temporal event feature. 
For temporal conditioning, we devise a temporal event feature and a novel conditioning technique named Block-FiLM.
T-Foley achieves superior performance in both objective and subjective evaluation metrics and generates Foley sound well-synchronized with the temporal events. 
Additionally, we showcase T-Foley's practical applications, particularly in scenarios involving vocal mimicry for temporal event control.
We show the demo on our companion website.\footnotemark
\end{abstract}


\section{Introduction}
\label{sec:intro}
Foley sound refers to human-created sound effects, such as footsteps or gunshots, to accentuate visual media. The significance of Foley sound lies in its ability to enhance the overall immersive experience for various forms of media \cite{dcase}. Foley sounds are usually created by Foley artists who record and produce required sounds manually, synchronized with the visual elements.

The advent of neural audio generation has presented an opportunity to automate and streamline the Foley sound creation process, reducing the time and effort required for sound production. 
To synthesize proper sounds from specific categories, early studies usually focused on single sound sources such as foosteps~\cite{footstep}, laughter~\cite{laughter}, and drum~\cite{drumgan, crash}. Subsequent research then further improved a model to be capable of generating multiple sound categories utilizing auto-regressive models~\cite{autoFoley, conditional_Foley_analogies, pixelsnail_vqvae}, Generative Adversarial Networks(GANs)~\cite{Foleygan, generating_visually}, or diffusion models~\cite{full}. Recently, it has been possible to generate holistic scene sounds solely based on a text description, even without pre-defined sound categories~\cite{diffsound, audioldm, audiogen}.

While prior research demonstrated the faithful sound synthesis by neural models for a target source, few focused on the timing and envelope of sound events. Precisely locating sound events is crucial for practical Foley sound synthesis. Some studies generated implicit temporal features from input videos during synthesis~\cite{autoFoley,Foleygan,generating_visually, variety_sound}. However, they lack explicit temporal event conditions for controllability and do not include quantitative analysis.

This research aims to address these challenges and produce realistic, timing-aligned Foley sound effects within a specific sound category. To the best of our knowledge, this is the first attempt to generate audio with explicit temporal event conditions.
Our contributions are the following: First, we propose \textbf{\textit{T-Foley}}, a \textbf{T}emporal-event-guided diffusion model with a conditioning sound class to generate high-quality \textbf{Foley} sound. For the temporal guidance, we introduce the temporal event feature to guide timing and envelope representation. To devise a conditioning method that reflects temporal informative condition, we introduce Block-FiLM, a modification of FiLM \cite{film} for block-wise affine transformation. 
Second, we conduct extensive experiments to validate the performance and provide a comparative analysis of results on temporal conditioning methods. A metric is proposed to measure the temporal fidelity quantitatively.
Lastly, we show the potential applications of T-Foley by demonstrating its performance on a human voice condition that mimics target sound events as an intuitive way to capture temporal event features. 
\footnotetext{\url{https://yoonjinxd.github.io/Event-guided_FSS_Demo.github.io}}

\begin{figure}[t]
    \centering \centerline{\includegraphics[width=7.5cm]{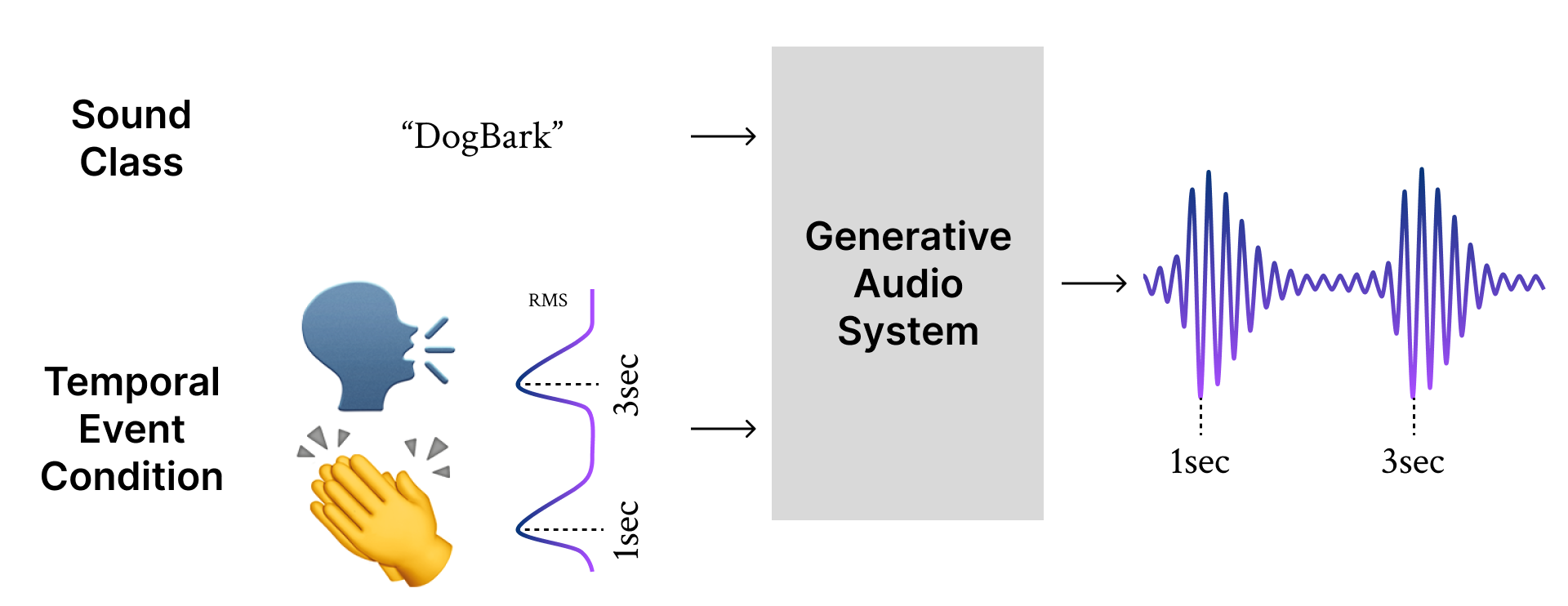}}
    \caption{Temporal-event-guided Foley sound synthesis.}
    \label{fig:task}
\end{figure}

\begin{figure*}[ht!]
    \centerline{\includegraphics[width=1.03\textwidth]{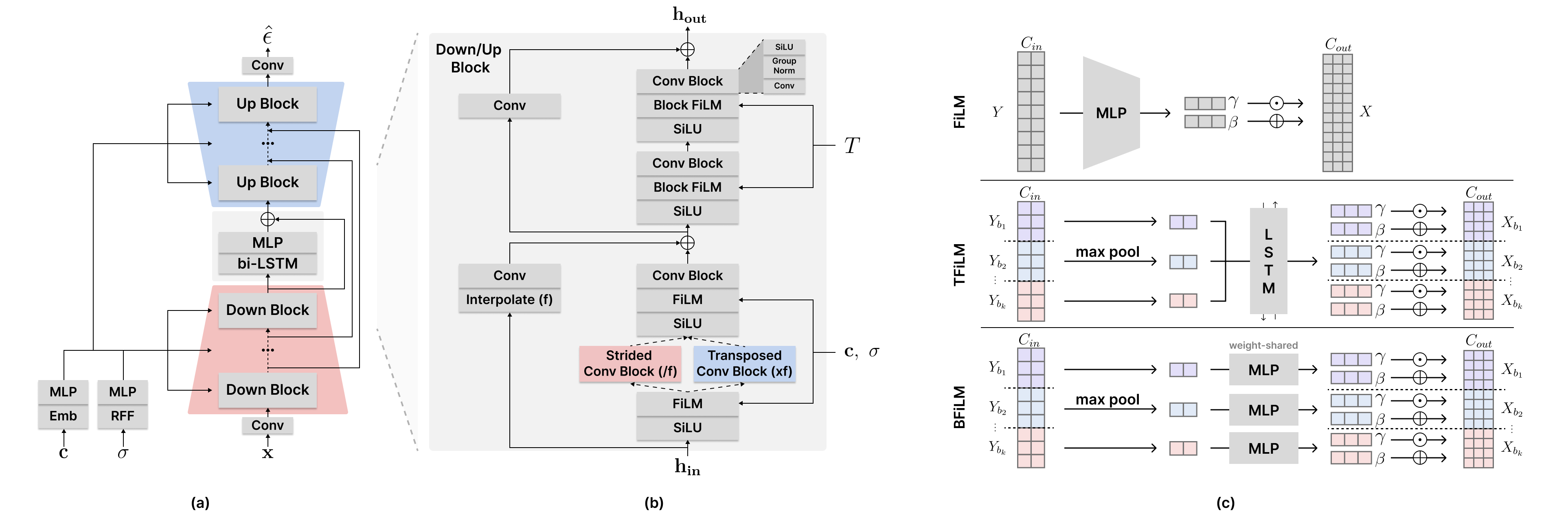}}
    \caption{(a) Overall architecture of the proposed model. ($\mathbf{c}$: sound class, $\sigma$: diffusion timestep, $T$: temporal event feature)
     (b) A detailed structure of a Down/Up sampling block. Each Down block includes strided convolutional layer at first while Up blocks exploit the transposed one. ($\mathbf{h_{in}}/\mathbf{h_{out}}$: latent features) (c) Comparison of FiLM, TFiLM, and the proposed BFiLM. ($Y$: conditioning input, $X$: input activation)} \label{fig:model}
\end{figure*}

\section{T-Foley} \label{sec:proposed}
As shown in Figure \ref{fig:task}, the model is designed to input the sound class category and temporal event condition, representing ``what'' and ``when'' the sound should be generated, respectively. 

\subsection{Overall Architecture}
The architecture of T-Foley is depicted in Figure \ref{fig:model}. Our U-Net design, with an auto-regressive bottleneck, is based on the advanced DAG~\cite{full} model, which produces high-resolution audio without a pretrained vocoder. Compared with CRASH~\cite{crash}, which is the first diffusion model to generate waveform from scratch, DAG contains a sequential module at the bottleneck of deeper U-Net architecture to address the problem of inconsistent timbre within a generated sample. 
To predict the noise of each diffusion timestep, the model first downsamples the noised signal $\mathbf{x}$ into the latent vector and passes it to the bidirectional LSTM to maintain the timbre consistency within a sample. The output of the bottleneck layer is resized by linear projections and finally upsampled into the prediction of noise $\hat{\epsilon}$. 
In each downsample and upsample block, convolution layers are conditioned on diffusion time step embedding $\sigma$ and class embedding $\mathbf{c}$ through FiLM, like in DAG, 
with the latter half employing Block-FiLM conditioned by temporal event feature $T$.
The model is trained end-to-end to minimize the continuous-time L2 loss on predicted noise as proposed in CRASH~\cite{crash}.

\subsection{Temporal Event Feature}\label{subsec:feature-representation}
As the primary objective of T-Foley is to generate audio upon a temporal event condition, the model needs to learn temporal information regarding the timing and envelope of sound events in time. This necessitates the appropriate conditioning temporal feature for the sound events. We used root-mean-square (RMS) of the waveform, which is widely used frame-level amplitude envelope features as below: for the i-th frame,  
\begin{equation}
    E_i(x)=\sqrt{{1\over W}\Sigma_{t=ih}^{ih+W}x^2(t)}
\label{eq.event_feature}
\end{equation}
where $x(t)\ (t\in[0,T])$ is the audio waveform, $W$ is a window size and $h$ is a hop size. In our experiment, we set $W=512$ and $h=128$.
We also considered power (the square of RMS) and onset/offset (the start and end of a particular sound event) as candidates. After a preliminary experiment, we decided to use RMS because there was no significant difference between RMS and power, and some categories (e.g., rain, sneeze) do not have definite onsets and offsets by the nature of the sounds but temporal patterns with varying intensities (i.e., envelope).

\subsection{Block FiLM}\label{subsec:block-film}
We propose Block FiLM as a neural module to condition the generation model on the temporal event feature. FiLM is one of the Conditional Batch Normalization (CBN) techniques to modulate the activations of individual feature maps or channels with affine transformation based on a conditioning input. It has been widely used in various tasks, including image synthesis, style transfer, and diffusion models for waveform generation \cite{wavegrad, crash, full}. \sloppy Mathematically, with a conditioning input $\mathbf{y}_{C_{in}, L_{in}}\in\mathrm{R}^{C_{in}\times L_{in}}$, where $C_{in}$ is the input channel size, and $L_{in}$ is the input length, the FiLM modulation can be expressed as follows:
$
    \text{FiLM}(\mathbf{x}, \mathbf{y}, \gamma, \beta) = \gamma\odot{\mathbf{x}} + \beta
$
where $\mathbf{x}$ represents the input activations, $\mathbf{c}$ is the conditioning input, $\gamma, \beta \in\mathrm{R}^{C_{out}}$ are normalizing parameters obtained as $\gamma, \beta = \text{MLP}(\mathbf{y})$. The $\odot$ symbol denotes the Hadamard product(element-wise matrix multiplication). 
  
Temporal FiLM (TFiLM) was proposed to overcome the limitation of FiLM in time-varying information in conditioning signals, which is crucial for processing audio signals. 
Given a sequential conditioning input $\mathbf{y_{C_{in}, L_{in}}}\in\mathrm{R}^{C_{in}\times L_{in}}$, where $C_{in}$ is the input channel size, and $L_{in}$ is the sequential length in the time domain, TFiLM firstly splits $\mathbf{y}$ into $N$ blocks. THe $i$-th block $Y_{b_i}\in\mathrm{R}^{C_{in}\times L_{in}/N}$ ($i=1,...,N$) is max-pooled in the time dimension as $Y^{\text{pool}}_{b_i}=\text{Max-Pool}_{1:L_{in}/N}(Y_{b_i})\in\mathrm{R}^{C_{in}}$ and followed by an RNN as sequential modeling to obtain normalizing parameters $(\gamma_i,\beta_i),h_i=\text{RNN}(Y^{\text{pool}}_{b_i};h_{i-1})$.    
Finally, a linear modulation is applied channel-wise according to the normalizing parameters:
\begin{equation}
    X^{\text{output}}_{b_i}=(\mathbf{1}_{L_{out}/N} \gamma_i^T)\odot X_{b_i}+(\mathbf{1}_{L_{out}/N} \beta_i^T)\in\mathrm{R}^{C_{out}\times L_{out}/N}
\end{equation}
where $\gamma_i, \beta_i\in\mathrm{R}^{C_{out}}$, $\mathbf{1}_{d}=[1,...,1]^T\in\mathrm{R}^d$. TFiLM was originally proposed for self-conditioning to modulate intermediate features. Therefore, $C_{in}=C_{out}$ and $L_{in}=L_{out}$ as $[X_{b_i}\text{'s}]^T=\mathbf{x}=\mathbf{y}=[Y_{b_i}\text{'s}]^T$. 
In this paper, we generalize it to the case where conditioning signal $\mathbf{y}_{C_{in}, L_{in}}$ and modulating signal $\mathbf{x}_{C_{out}, L_{out}}$ are different.
However, incorporating TFiLM in every conditioning layer can significantly increase computational complexity.

Block FiLM (BFiLM) is a simplified version of TFiLM motivated by the characteristics of RMS.
The temporal events embedded in sequential information of RMS are weakly dependent (e.g., a sound event at t=1.3sec does not affect another sound event at t=3sec.)
Therefore, we suggest adopting block-wise transformations from TFiLM by replacing the unnecessary sequential modeling layer with a simple MLP layer as the following:
\begin{equation}
    (\gamma_i,\beta_i)=\text{MLP}(Y^{\text{pool}}_{b_i}).
\end{equation}
T-Foley is designed under the assumption that the LSTM layer located at the bottleneck of U-Net architecture is capable of handling the sequence modeling among the blocks.We demonstrate the performance and efficiency of BFiLM in Section \ref{subsec:comp_films}.

\begin{table}[]
\centering
\resizebox{\columnwidth}{!}{%
\begin{tabular}{@{}lcc|ccccc@{}}
\toprule
Model &  \#params$\downarrow$ & infer.t $\downarrow$  & E-L1$\downarrow$   & FAD-P$\downarrow$   & FAD-V$\downarrow$  & IS$\uparrow$  \\ \midrule
Real data & - & - & 0.0 &        22.81 & 4.06 & 2.18  \\ \midrule
DAG\cite{full} (w/o temp. c.)& 87M & 12s &    0.2212 & 53.94 & 36.10 & 1.46  \\
T-Foley (FiLM\cite{film}) & 83M & \textbf{6.3s} &    0.0772 & 54.59 & 36.06 & \textbf{1.94} \\
T-Foley (TFiLM\cite{tfilm}) & 101M & 13s &       0.0469 & 49.44 & 36.10 & 1.74 \\
T-Foley (BFiLM) & \textbf{74M} & 9.5s &      \textbf{0.0367}  & \textbf{41.59} & 36.09 & 1.79  \\ 
\bottomrule
\end{tabular}
}
\caption{Objective evaluations of reproduced DAG(which lacks temporal condition) and our T-Foley conditioned by FiLM, TFiLM and BFiLM.
(\#params: Number of trainable parameters, infer.t: Approximate inference time for predicting 1 sample, E-L1: Event-L1 Distance, FAD-P, and FAD-V: FADs based on PANNs and VGGish, IS: Inception Score.)}
\label{tab:result_quan}
\end{table}


\section{Experimental Setup}
\label{sec:experiments}

\subsection{Datasets}
We utilized the Foley sound dataset from the Foley sound synthesis task of 2023 DCASE challenge~\cite{dcase}. The given dataset comprises approximately 5k class-labeled sound samples (5.4 hours) sourced from three different datasets. The dataset covers 7 Foley sound classes: \textit{DogBark, Footstep, GunShot, Keyboard, MovingMotorVehicle, Rain, Sneeze\_Cough}. All audio samples were constructed in mono 16-bit 22,050 Hz format, each lasting for 4 seconds. We used approximately 95\% of the development dataset for training and 5\% for validation in this work.

While controllable Foley sound synthesis holds great potential, expressing desired temporal events can be non-trivial for users.
For more intuitive conditioning, we used human voices that mimic Foley sounds as a reference to extract temporal event conditions. In particular, we used subsets of two vocal mimicking datasets paired with the original target sounds: Vocal Imitation Set \cite{vocalimitation} and VocalSketch \cite{vocalsketch}.  
We adjusted the duration of each audio sample representing the 6 sound classes (excluding Sneeze\_Cough) to match the training data.

\subsection{Experimental Details}
We train our model to estimate the reparameterized score function of a normal transition kernel with variance-preserving cosine scheduling as proposed in \cite{crash}. 
For conditional sampling, we employ DDPM-like discretization of SDE \cite{vdm} with classifier-free guidance \cite{classifierfree}. During the 500-epoch of training, we randomly dropped the conditions $p=0.1$ for training in the unconditional scenario.

\subsection{Objective Evaluation}
Objective Evaluation of audio generation models relies on three metrics.
FAD and IS measure that the generated sounds align with the given class condition and their diversity. 
For FAD, we exploit two classifiers from VGGish model\cite{vggish}(FAD-V, 16kHz) and PANNs\cite{panns}(FAD-P, 32kHz).
IS also utilizes PANNs.
To verify the effectiveness of the temporal condition, we exploit \textbf{\textit{Event-L1 Distance (E-L1)}}. E-L1 assesses how well the generated sounds adhere to the given temporal event condition. We employed the L1 distance between the event timing feature of the target sample and the corresponding generated sample as follows:
\begin{equation}
    E\text{-}L1 = {1\over k}\displaystyle\Sigma_{i=1}^k ||E_i-\hat E_i||
\label{eq.LCE}
\end{equation}
where $E_i$ is the ground-truth event feature of $i$-th frame, and $\hat E_i$ is the predicted one.
We average the class-wise scores in each case.

\subsection{Subjective Evaluation}
A total of 23 participants conducted subjective evaluations, providing assessments for two types of generated samples. These types include: 1) samples generated with temporal conditions from Foley sound test dataset, and 2) samples generated with temporal conditions from human vocals mimicking Foley sounds.
Participants rated the generated samples on a scale from 1 to 5, in 0.5-point increments based on three criteria:
\textit{Temporal Fidelity (TF)} for alignment of the generated samples with the temporal event condition of the target sample, \textit{Category Fidelity (CF)} for suitability to the given category, and \textit{Audio Quality (AQ)} for the overall quality of the generated sample. We report the Mean Opinion Score (MOS) derived from the participants' ratings.

\begin{table}[t]
\centering
\resizebox{\columnwidth}{!}{%
\begin{tabular}{@{}lcccc@{}}
\toprule
Model & Category Fidelity$\uparrow$  &  Temporal Fidelity$\uparrow$  & Audio Quality$\uparrow$   \\ \midrule
FiLM\cite{film} & 3.85($\pm$0.12) & 4.11($\pm$0.10) & 3.28($\pm$0.11) \\
TFiLM\cite{tfilm} & 4.02($\pm$0.11) & 4.00($\pm$0.13)& 3.75($\pm$0.11) \\
BFiLM   & \textbf{4.22}($\pm$0.11) & \textbf{4.41}($\pm$0.09) & \textbf{4.06}($\pm$0.10)  \\ 
\bottomrule
\end{tabular}%
}
\caption{Comparison of Mean Opinion Scores(MOS). Mean and 95\% confidence interval are reported.}
\label{tab:result_qual}
\end{table}
\section{Results}

\subsection{Temporal Event Conditioning Methods}\label{subsec:comp_films}
We compare the performance of different conditioning methods for the temporal event condition. Table \ref{tab:result_quan} and \ref{tab:result_qual} present the objective scores and MOS (Mean Opinion Score) from subjective evaluation. Overall, TFiLM and BFiLM, which consider the temporal aspect of the event condition, received higher scores in most of the objective and subjective metrics. Notably, BFiLM demonstrated markedly superior performance in most of the results, particularly achieving improvements with approximately 0.7 times fewer parameters and less inference time compared to TFiLM. These results validate the hypothesis on the efficiency of BFiLM (Section\ref{subsec:block-film}).
FiLM may have high IS value due to generating diverse low-quality audio different from ground-truths. In other experiments, we only use BFiLM.

\subsection{Effect of Block Numbers}\label{subsec:blocks}
\begin{figure}[]
    \centerline{\includegraphics[width=8.cm]{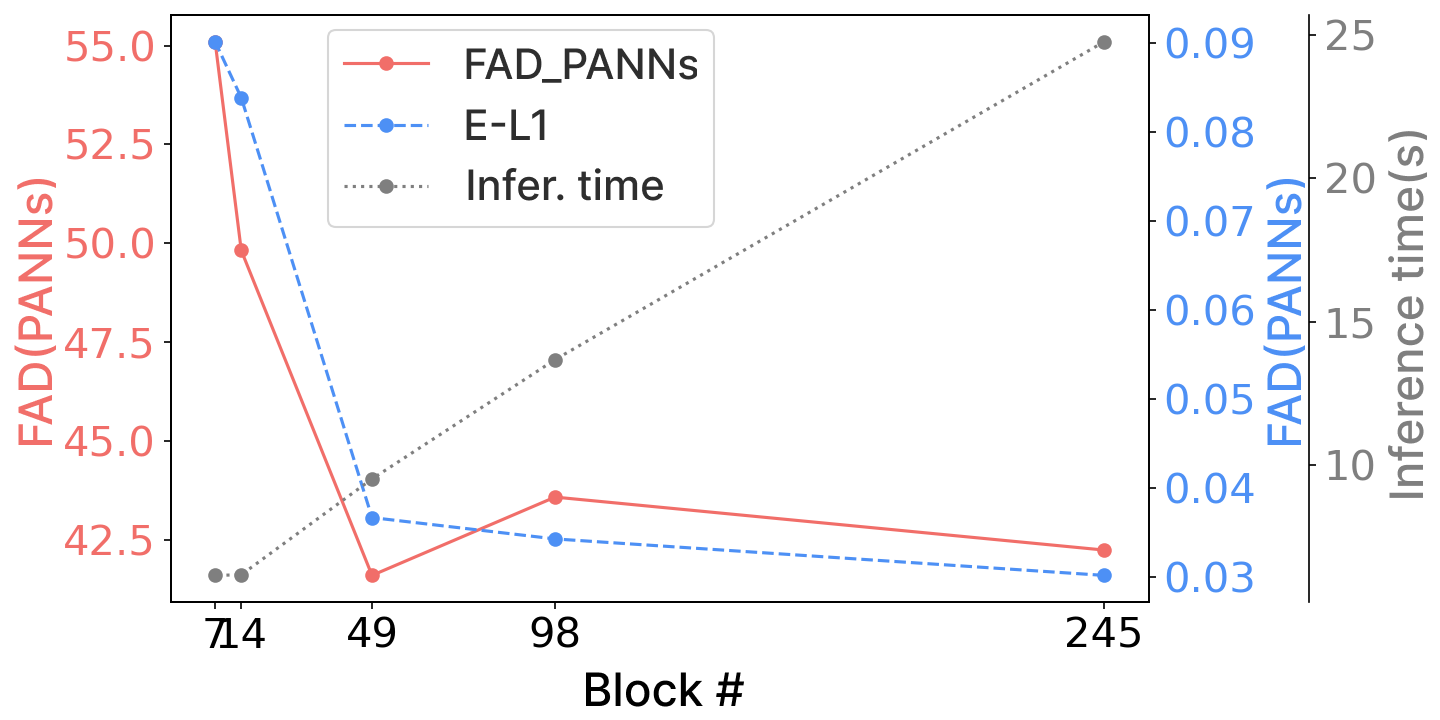}}
    \caption{Tradeoff between performance(E-L1, FAD-P) and efficiency(inference time) among different number of blocks.} \label{fig:tradeoff}
\end{figure}

Block Number $N$ in Section \ref{subsec:block-film} is an important hyperparameter that influences the resolution of the condition. Fewer blocks lead to sparser and smoother conditional information in the temporal axis. We compare the performance of different block numbers in Figure \ref{fig:tradeoff}. For accuracy, E-L1 decreases as there are more blocks as expected. FAD-P also decreases, without significant difference among 49, 98, and 245. In terms of efficiency, inference time increases along with the block number, as it requires more computation. Considering the tradeoff between accuracy and efficiency, we stick to 49 blocks in other experiments.

\begin{table}[]
\centering
\resizebox{6.cm}{!}{
\begin{tabular}{@{}lcccc@{}}
\toprule
\multirow{2}{*}{Block \#} & \multicolumn{2}{c}{Vocal Imitation} & \multicolumn{2}{c}{VocalSketch} \\
    & E-L1$\downarrow$   & FAD-P$\downarrow$ & E-L1$\downarrow$   & FAD-P$\downarrow$ \\ \midrule
245 & 0.0228 & 74.94 & 0.0186 & 59.33 \\
98  & 0.0300 & 69.13 & 0.0274 & 56.51 \\
49  & 0.0306 & 64.55 & 0.0299 & 49.62 \\
14  & 0.0764 & 59.56 & 0.0635 & 58.93 \\
7   & 0.0935 & 66.23 & 0.0914 & 60.41 \\ \bottomrule
\end{tabular}
}
\caption{Evaluation on Vocal Mimicking Datasets}
\label{tab:vocal}
\end{table}

\subsection{Evaluation on Vocal Mimicking Datasets}
Table \ref{tab:vocal} summarizes the results among different block numbers, showing comparable performance in vocal sound. E-L1 decreases for larger block numbers, which is consistent with Section \ref{subsec:blocks}. On the other hand, FAD-P is the lowest around 14, 49 number of blocks. This result may arise from the differences between Foley sound and vocal, as the two sounds exhibit distinct RMS curves due to discrepancies in timbre and energy envelope. Therefore, choosing the right block number is crucial for adjusting the smoothness of RMS to match the conditioning feature with the target sound's characteristics. MOS for model with $N=49$ was measured as CF$=4.41(\pm 0.09)$, TF$=4.40(\pm 0.10)$, and AQ$=4.34(\pm 0.08)$, providing a competitive result with Table \ref{tab:result_qual}.

\begin{figure}[t]
    \centerline{\includegraphics[width=6.cm]{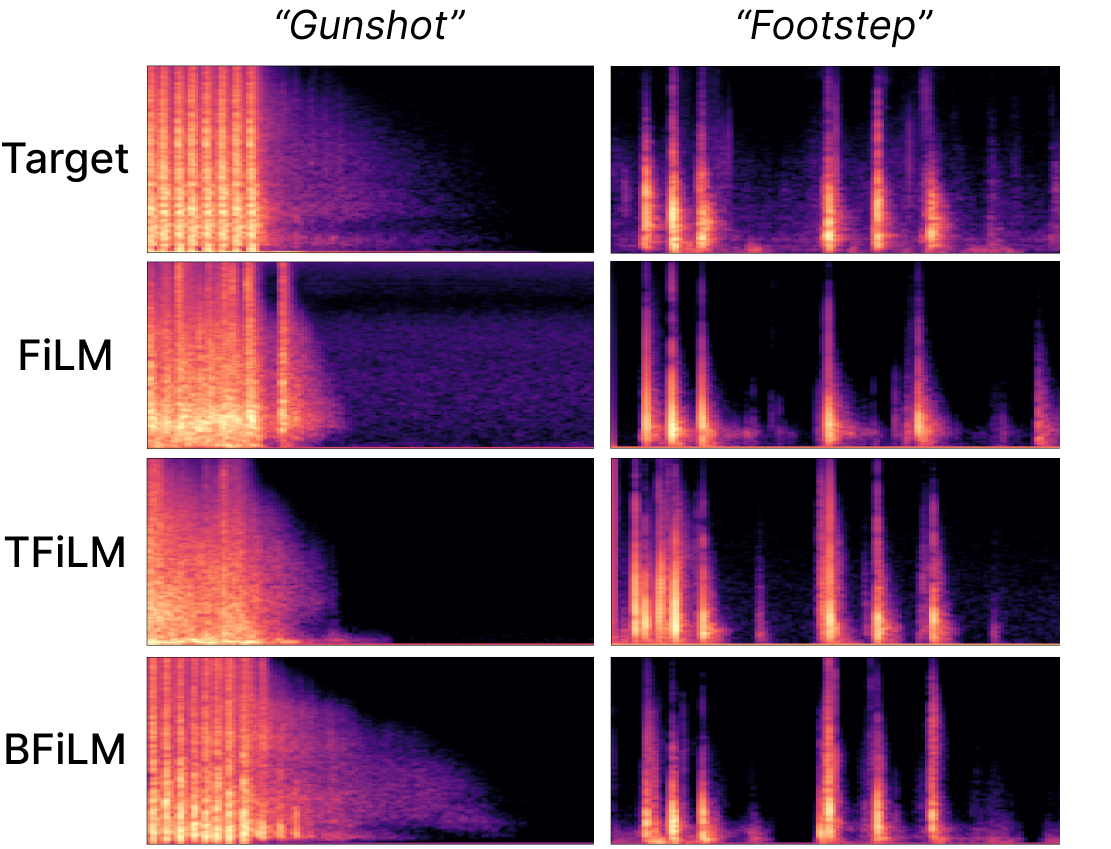}}
    \caption{The first row shows the control sounds used to extract the target event feature. Subsequent rows show three classes of Foley sounds generated with different conditioning blocks (FiLM, TFiLM, and BFiLM), all represented as mel-spectrograms.}
    \label{fig:sample_spectrograms}
\end{figure}

\begin{figure}[t]
    \centerline{\includegraphics[width=6.cm]{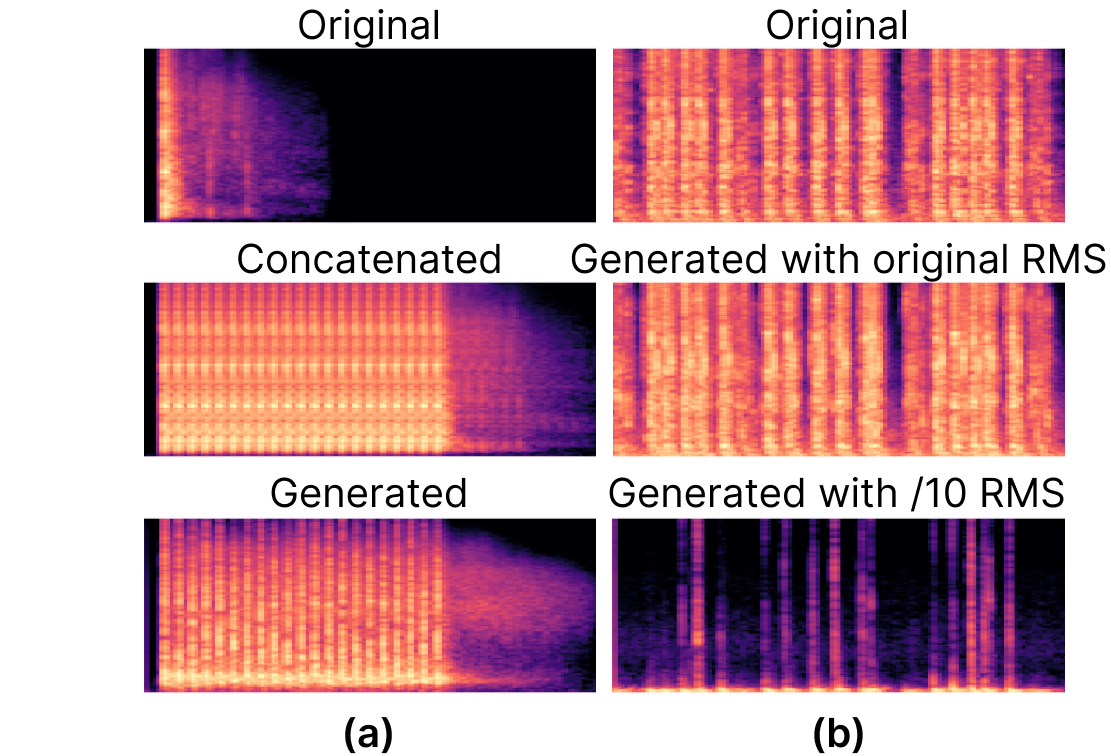}}
    \caption{(a) Comparing manually synthesized consecutive gunshot sounds with sounds generated through temporal event feature. (b) Generated sounds with the original temporal event features and those with a reduced gain by 10.} \label{fig:control_samples}
\end{figure}

\subsection{Case Study} 
To showcase the performance and usability of T-Foley, we present two case studies. Firstly, we compare the output of our proposed BFiLM method with that of the FiLM and TFiLM methodologies. In Figure \ref{fig:sample_spectrograms}, BFiLM exhibits the highest alignment with the mel-spectrogram of the target conditioning sound. 
Both FiLM and TFiLM generate unclear and undesirable sound events in the Footstep sound class. For Gunshot, only BFiLM seems to reflect the sustain and decay of the attack in sound well.

Furthermore, the generation of Foley sounds using temporal event conditions yields considerably more realistic results when compared to manual Foley sound manipulation. We exemplify two specific scenarios in Figure \ref{fig:control_samples}. The first scenario involves consecutive machine gunshots.
Manually adjusting and copying individual gunshot sound snippets can result in an unnatural audio sequence. Conversely, employing T-Foley to concatenate temporal event conditions leads to a seamless and lifelike sound. The second is a key-typing sound with two contrasted examples: typing vigorously on a typewriter and softly pressing keys on a plastic keyboard.
T-Foley can generate these two sounds by adjusting the gain of the temporal event feature. This indicates that the level of temporal event feature controls not only the amplitude but also the timbre texture.  
In addition to the showcased examples, you can explore a wider array of demonstration samples and real-world scenarios (such as claps and voices) on our accompanying website introduced in the abstract.

\section{Conclusion}
\label{sec:conclusion}
This study presents T-Foley, a Foley sound generation system addressing controllability in the temporal domain. By introducing the temporal event feature and the E-L1 metric, we show that Block-FiLM, our proposed conditioning method, is effective and powerful in terms of quality and efficiency. We also demonstrate the performance of T-Foley on vocal mimicking datasets to claim its power in usability. 
Foley sounds can be broadly categorized into two groups: transient event-based sounds and continuous ambient sounds, each with distinct temporal characteristics. While we have not specifically differentiated between the two, we believe acknowledging and addressing this distinction could enhance performance.

\vfill \pagebreak
\bibliographystyle{IEEEbib}
\bibliography{strings,refs}

\end{document}